\documentclass[aps,prd,amsmath,amssymb,twocolumn,groupedaddress]{revtex4-2}  
\usepackage{graphicx}
\usepackage{hyperref}
\usepackage{natbib}
\usepackage{xcolor}
\usepackage{mathtools}

\definecolor{dbl}{rgb}{0, 0, 0.9}
\hypersetup{
  bookmarksopen=true,     
  colorlinks=true,        
  linkcolor=dbl,     
  citecolor=dbl,      
  filecolor=dbl,      
  urlcolor=dbl        
}

\usepackage{color}

\begin{document}

\title{Bright Cosmic-Ray Source as a Solution to Auger-TA Tensions}

\author{Alexander Korochkin}
\email{alexander.korochkin@ulb.be}
\affiliation{
 Université Libre de Bruxelles, CP225 Boulevard du Triomphe, 1050 Brussels, Belgium
}
\author{Dmitri Semikoz}
\affiliation{
 Université de Paris Cite, CNRS, Astroparticule et Cosmologie, F-75013 Paris, France
}
\author{Peter Tinyakov}
\affiliation{
 Université Libre de Bruxelles, CP225 Boulevard du Triomphe, 1050 Brussels, Belgium
}
\date{\today}

\begin{abstract}
The ultra-high-energy cosmic ray (UHECR) spectra measured by the Pierre Auger Observatory (Auger) and the Telescope Array (TA) agree very well below $10^{19.5}$~eV but differ significantly at higher energies. We show that these differences can be explained by a single nearby source superimposed on a nearly isotropic background. Taking into account deflections in Galactic and extragalactic magnetic fields, such a source can account for the excess in the TA spectrum without producing excessive anisotropy. The required hard spectrum of the source and intermediate-mass composition are consistent with previous fits of the Auger-only spectrum and composition. This scenario offers several additional advantages: (i) the source produces a broad excess partially overlapping the TA hotspots, suggesting their possible explanation; (ii) without additional tuning, it reproduces the $\sim90^\circ$ shift in dipole direction observed between the Auger-only and combined Auger–TA analyses; and (iii) the best-fit position of the source lies near M82, the brightest nearby starburst galaxy, making it a plausible source of the UHECR.
\end{abstract}

\maketitle

\section{Introduction}
The energy spectra of ultra-high-energy cosmic rays (UHECRs) measured by the Pierre Auger Observatory (Auger)~\cite{PierreAuger:2015eyc} and the Telescope Array (TA)~\cite{Tokuno:2012mi,TelescopeArray:2012uws} agree remarkably well over the range $10^{18}$--$10^{19.5}$\,eV, once a simple 9-11\% energy rescaling is applied between the two experiments~\cite{TelescopeArray:2021zox,Auger_TA_ICRC2025}. This shift lies well within their systematic uncertainties (14\% for Auger, 21\% for TA). Furthermore, if both experiments use the same fluorescence yield model and apply the same treatment of the invisible energy correction $E_\mathrm{inv}$, the required rescaling decreases to only 4\%~\cite{Auger_TA_ICRC2025}. This energy range includes two features --- the so-called ``ankle'' at $E \approx 10^{18.75}$\,eV and the ``instep'' at $E \approx 10^{19.15}$\,eV --- both coinciding in position and shape after rescaling. Surprisingly, at higher energies, $E > 10^{19.5}$\,eV, the spectra diverge (Fig.~\ref{fig:spec_TA_Auger}), with a discrepancy of $\sim 8\sigma$~\cite{TA_8sigma}, ruling out a purely statistical origin.

Energy-dependent systematics in the spectra reconstruction could, in principle, account for this discrepancy~\cite{Plotko:2022urd}: a low-energy calibration shift may not be correct in the highest-energy region. However, neither of the two experiments has found indications that this might be the case~\cite{Auger_TA_ICRC2025}. In addition, the Auger and TA spectra have been compared and found consistent in the region of overlapping exposures~\cite{TA_8sigma}. 

This suggests that the discrepancy at the highest energies is of a physical origin: it may be due to Auger and TA different fields of view (Southern vs. Northern Hemisphere), and the UHECR spectrum may vary across them. Supporting this idea, the TA collaboration reported evidence that the UHECR spectrum is declination-dependent within the TA field of view~\cite{TA_declin}. An analogous test by the Auger collaboration found no such dependence within their field of view (FoV)~\cite{PierreAuger:2025hnw}.

If the spectral differences are physical, they must originate from anisotropic distribution of sources across the sky. In that case, other anisotropies would likely appear and must also be consistent with the data. Yet the observed UHECR flux is close to isotropic: the only confirmed deviation is a dipole modulation with an amplitude of $\sim 6$\%, detected in the Auger data for $E > 8$~EeV~\cite{PierreAuger:2017pzq, 2024ApJ...976...48A}. While at these energies the dipole direction is well reproduced by sources following the large-scale structure (LSS) \cite{Allard:2021ioh,Bister:2023icg,Bister:2024ocm}, at higher energies $E>32$~EeV the situation is more complicated. When inferred from Auger-only data, the dipole direction is close to the low-energy one and is still consistent with LSS sources. However, from the joint Auger and TA data a significantly different direction is found \cite{TelescopeArray:2021ygq, PierreAuger:2023mvf, TelescopeArray:2025yvu}, which is in tension with the LSS predictions. 

In addition to the dipole anisotropy, at $E \gtrsim \text{(a few)} \times 10^{19}$\,eV, several localized excesses of events (``hotspots'') have been reported \cite{PierreAuger:2010ofq,PierreAuger:2024hrj, TA_hotspot,TA_warmspot}, though with statistical significance still insufficient for firm conclusions. The source distribution which explains the spectral differences must also avoid overproducing these medium-scale anisotropies. 

The purpose of this paper is to show that it is possible, in principle, to reconcile  apparently conflicting requirements of general isotropy and spectral difference between Auger and TA by assuming that the UHECR flux at the highest energies consists of two components: a smooth nearly-isotropic part consisting of a monopole and dipole of amplitude $6\%$ as measured by Auger, plus a contribution from a {\em single} close source with a hard spectrum placed in the Northern hemisphere. This toy model may be considered an approximation to the situation in which the smooth flux is produced by numerous sources following the LSS, while the additional flux comes from an anomalously close source. This is a conservative approach: clearly, if the data can be fitted with a single source on top a smooth background, they can be fitted even better with a few sources of a similar total magnitude. 

There have been several attempts to identify UHECR sources by cross-correlating cosmic-ray arrival directions with various classes of astrophysical objects without accounting for deflections in the Galactic magnetic field (GMF) (see, e.g., \cite{Tinyakov:2001nr,PierreAuger:2007edx,Kotera:2008ae,PierreAuger:2012mqu,Oikonomou:2012ef,PierreAuger:2018qvk,TelescopeArray:2018qwc,Kim:2019eib}). However, these deflections are typically large in the case of nuclei \cite{Unger:2023lob,Korochkin:2025ugg} and therefore cannot be neglected~\cite{Higuchi:2022xiv}.

A key ingredient of our toy model is the \texttt{KST24} Galactic magnetic field  configuration proposed in \cite{Korochkin:2024yit} (for an alternative models see \cite{Han:1999vi,Han:2006ci,Pshirkov:2011um,JF_GMF_1,Shaw:2022lqd,Dickey:2022fji,Unger:2023lob}). 
We find that with the best-fit values of the field parameters and intermediate particle charge, UHECRs may have just the right magnitude of deflections to avoid too strong intermediate-scale anisotropy while still explaining the Auger-TA spectral difference. As a bonus, the model reproduces qualitatively several other observed features of the UHECR flux: (i) due to a relatively strong magnetic field in the direction of the outer Galaxy (the Fan region), the source produces a mild and wide excess roughly in the direction of the TA hot spots; (ii) the direction of the dipole at high energies $E>32$~EeV shifts close to the position measured at these energies in the combined Auger-TA data set. Interestingly, the required parameters of the source --- its spectral index, cutoff energy and injection mass composition --- are consistent with ones derived in Auger-only study~\cite{2023PhRvD.107j3045E} for the bulk of the UHECR sources. 

The rest of this paper is organized as follows. Section~\ref{sec:setup} outlines the details of the analysis and the data used. The results of the fits, particularly the required parameters and position of the nearby source, are presented in Section~\ref{sec:results}. Finally, in Section~\ref{sec:summary} we discuss the implications of our finding.

\section{UHECR flux Model and data analysis}
\label{sec:setup}

The UHECR flux in our toy model consists of the background with the dipolar modulation and a contribution from a single source. The parameters of the background, the position of the source on the sky, its chemical composition and spectrum as described in detail below, are considered free parameters which we fit to the UHECR data, aiming to reproduce the Auger-TA spectral differences without introducing excessive anisotropy. 

We parametrize the background $I_\mathrm{bkg}(E)$ above $10^{18.8}$~eV using a broken power-law with two break points, as this form has been shown to provide a good fit to the Auger spectrum~\cite{PierreAuger:2025hnw}. Additionally, we incorporate the dipole anisotropy as measured by Auger. We keep its amplitude and direction fixed to the observed values, assuming it is produced by the population of background sources that follow the LSS.

To fit the model, we used the latest publicly available spectra from TA~\cite{TA_8sigma} and Auger~\cite{PierreAuger:2025hnw}. The TA spectrum comprises the  data collected between May 11, 2008, and May 10, 2022 for the zenith angles less than $55^\circ$. The Auger spectrum is based on the data recorded between January 1, 2004 and December 31, 2022 for the zenith angles less than $80^\circ$. The TA and Auger Joint Spectrum Working Group has found in ref.~\cite{Auger_TA_ICRC2025} that, to align the spectra of the two experiments at energies around the ankle at  $E=10^{18.7}$~eV, the relative energy shift of 11\% should be applied (e.g., the Auger energy scale raised by 5.5\% and TA -- lowered by 5.5\%). 

Since we are using spectra based on different datasets~\cite{TA_8sigma, PierreAuger:2025hnw} and fit the model at higher energies above $E=10^{18.8}$~eV, we recalibrate the spectra in the energy range $10^{18.8} - 10^{19.5}$~eV. The lower limit is set by the lowest-energy data point available in the TA spectrum~\cite{PierreAuger:2025hnw}, and the upper limit corresponds to the energy at which the TA bump appears. The energy calibration also accounts for the Auger-measured dipole anisotropy, extrapolated to the TA FoV.

\begin{figure}
    \centering
    \includegraphics[width=1.03\linewidth]{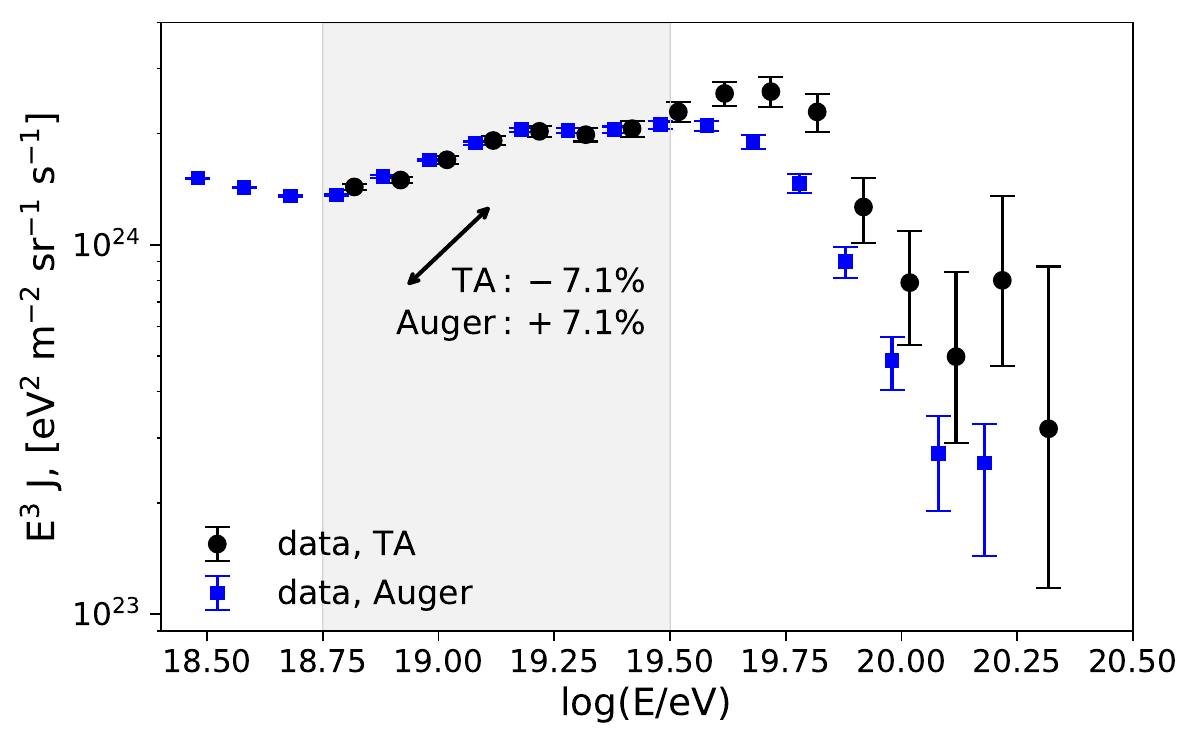}
    \caption{TA~\cite{TA_8sigma} and Auger~\cite{PierreAuger:2025hnw} spectra shifted to the best-fit common energy scale in the energy range from $10^{18.75}$~eV to $10^{19.5}$~eV (highlighted with the gray shading).
    \label{fig:spec_TA_Auger}}
\end{figure}

We find a best-fit value of 7.1\% for the symmetric energy-scale shift instead of 5.5\%, which is still well within the systematic uncertainties of both experiments. This energy-scale shift is kept fixed in the subsequent analysis. Symmetrically shifted energy spectra of Auger and TA are shown in Fig.\ref{fig:spec_TA_Auger}. The spectra agree below $10^{19.5}$ eV, with TA points lying a few percent lower due to their field of view being near the Auger dipole minimum. To the contrary, above this energy the TA spectrum is systematically higher.

For the source spectrum, we assume a power-law with an exponential cutoff in rigidity $R$:
\begin{equation}
	\frac{\mathrm{d}N}{\mathrm{d}R} \equiv F(R) = C \left(\frac{R}{R_0}\right)^{-\gamma} \exp\left[-\left(\frac{R}{R_\mathrm{cut}}\right)^\beta\right],
\end{equation}
where $R_0 = 10^{18}$~V. The normalization $C$, spectral index $\gamma$, cutoff rigidity $R_\mathrm{cut}$ and cutoff steepness $\beta$ are all treated as free parameters, with their values determined during the fit. The source is assumed to follow a Peters cycle~\cite{Peters:1961mxb}, therefore the total spectrum is given by
\begin{equation}
	F_\mathrm{src}(E) = \sum_i \frac{f_i}{Z_i}\,F(E/Z_i),
\end{equation}
where $f_i$ are the elemental injection fractions and $Z_i$ are the corresponding charges. The fractions $f_i = \{f_\mathrm{H}, f_\mathrm{He}, f_\mathrm{C}, f_\mathrm{Si}, f_\mathrm{Fe}\}$ satisfying $\sum f_i=1$ represent five effective mass groups in the UHECR injection composition ($ \prescript{1}{}{\mathrm{H}}, \prescript{4}{}{\mathrm{He}}, \prescript{12}{}{\mathrm{C}}, \prescript{28}{}{\mathrm{Si}}, \prescript{56}{}{\mathrm{Fe}}$) and are treated as free parameters in the fit. Normalized by $Z_i$, fractions $f_i$ represent the relative flux ratios at the same rigidity. As the source is assumed to be nearby, the UHECR interactions on the way from the source to the observer were neglected.

The fitting procedure is based on the maximum likelihood method. Technically, we minimize the quantity  $\chi^2 = -2\log \mathcal{L}$, where the total likelihood consists of two terms
\begin{equation}
	\log \mathcal{L} = \log \mathcal{L}_\mathrm{spec} + \log \mathcal{L}_\mathrm{ani}.
\end{equation}
The first term, $\log \mathcal{L}_\mathrm{spec}$, is the Poisson likelihood for the energy spectra of TA and Auger
\begin{equation}
	\log \mathcal{L}_\mathrm{spec} = -\sum_i \left(\mu_i - n_i + n_i \log\frac{n_i}{\mu_i}\right).
\end{equation}
Here the sum runs over all TA and Auger energy bins with \mbox{$E>10^{18.8}$~eV}, and $n_i$ and $\mu_i$ are the observed and expected number of events, respectively.

The model prediction $\mu$ is computed in two steps. First, we construct a skymap of the UHECR intensity as the sum of the background and the source image at Earth, i.e., after propagation through the GMF
\begin{equation}
	\frac{\mathrm{d}N}{\mathrm{d}E \mathrm{d}\Omega} = I(E, \Omega) = I_\mathrm{bkg}(E, \Omega) + I_\mathrm{src}(E, \Omega).
\end{equation}
For the coherent GMF, we adopt the \texttt{KST24} model~\cite{Korochkin:2024yit}, combined with the corrected model for the turbulent component from~\cite{JF_GMF_2}. To account for possible additional deflections by magnetic fields, we also perform the calculations with an extra angular smearing of $\theta_{19}^\mathrm{IGMF} =  \{10^\circ, 20^\circ, 30^\circ\}$ at $R=10^{19}$~V. The smearing at other rigidities is scaled proportionally. This additional smearing may arise, for example, from a strong intergalactic magnetic field (IGMF) with $n$G strength between the source and the observer, or from an extended magnetic halo of the Milky Way. Here we neglect any flux amplification or deamplification by the IGMF~\cite{Dolgikh:2022yep,Dolgikh:2023bke,Dolgikh:2025bac}, assuming its correlation length is sufficiently small.

Next, the resulting skymap is convolved with the directional exposure $A(\Omega)$ of either TA or Auger to obtain the expected number of events
\begin{equation}
	\mu_i = \int\limits_{E_i}^{E_i + \mathrm{d}E_i} \mathrm{d}E \int\limits_\mathrm{FoV} \mathrm{d}\Omega \,\,I(E, \Omega) A(\Omega).
\end{equation} 
The energy integral is taken over the $i$-th energy bin, while the angular integral covers the field of view (FoV) of TA or Auger. Since we focus on the highest energies where both detectors operate at full efficiency, we adopt purely geometrical exposure for both experiments.

The second term of the total likelihood, $\log \mathcal{L}_\mathrm{ani}$, represents a penalty for overproducing medium-scale anisotropy. A sufficiently bright spot --- an image of a point source after processing through magnetic field --- would likely have already been detected by either TA or Auger; thus, its absence in the data places an upper bound on the source brightness as a function of the angular size of its image. To implement this constraint, we evaluate the model predictions using the Li-Ma method. During each fit iteration, the model intensity map $I(E, \Omega)$ is computed on a \texttt{NSIDE}=32 HEALPix grid. The Li-Ma significance is then calculated for the cumulative skymap
\begin{equation}
    I(\Omega) = \int\limits_{E_\mathrm{thr}}^{\infty} \mathrm{d}E \,\,I(E, \Omega) A(\Omega),
\end{equation}
which represents the probability distribution for detecting events above $E_\mathrm{thr}$. This allows us to directly apply the standard TA and Auger intermediate-scale hotspot search methodology.

We normalize $I(\Omega)$ to the total number of observed events and perform the Li-Ma analysis for each of the two experiments for the energy thresholds $E_\mathrm{thr} = \{19.4, 19.5, 19.6, 19.75, 19.8\}$ and sampling circles of radii $\Theta = \{15^\circ, 20^\circ, 25^\circ, 30^\circ, 35^\circ\}$. The centers of the circles are taken in the centers of the pixels of $I(\Omega)$. As a result, the Li-Ma significance maps $I_\mathrm{\sigma}(\Omega)$ are obtained. Finally, we use $I_\mathrm{\sigma}(\Omega)$ to compute the penalty term $\log \mathcal{L}_\mathrm{ani}$, which effectively excludes all source configurations producing hotspots with significance exceeding $5\sigma$ in the TA FoV and $2.5\sigma$ in the Auger FoV. Otherwise, the penalty term is zero.

\section{Results}
\label{sec:results}
\begin{figure*}
    \centering
    \includegraphics[width=0.49\linewidth]{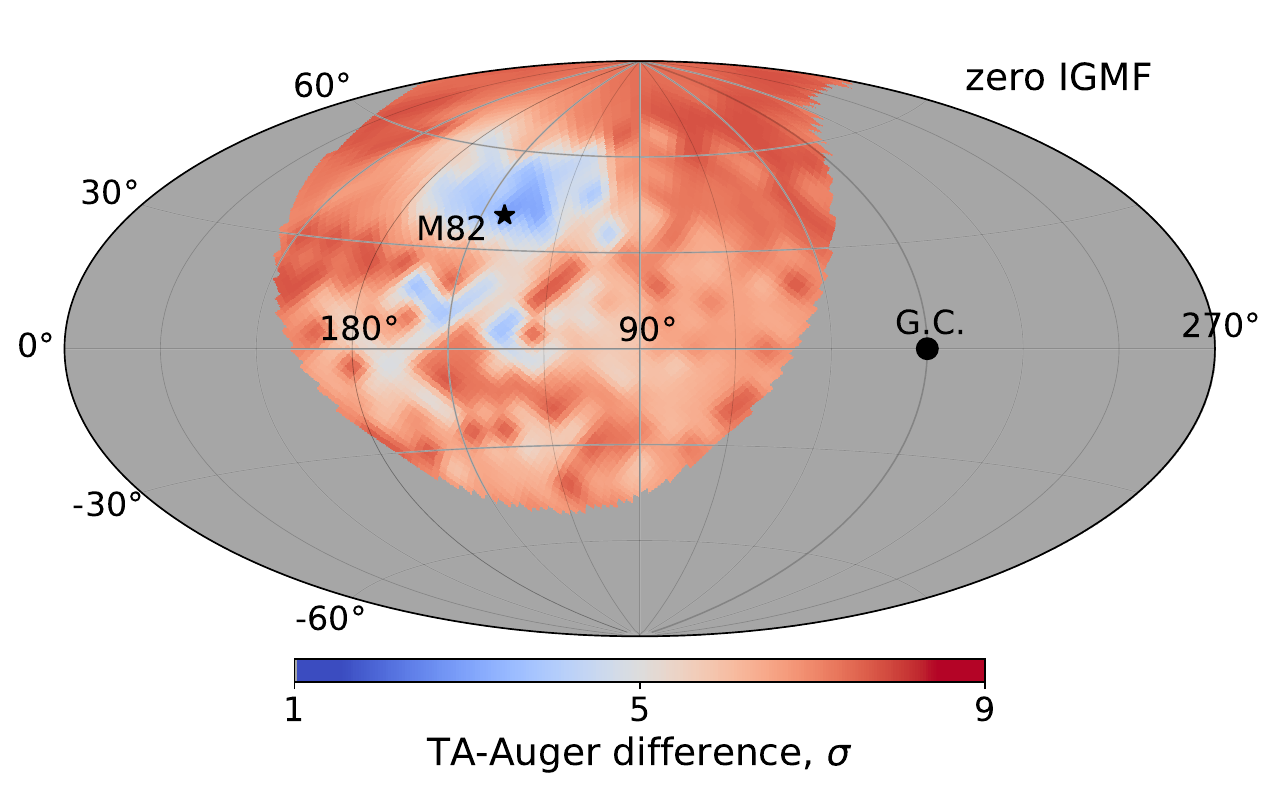}  
    \includegraphics[width=0.49\linewidth]{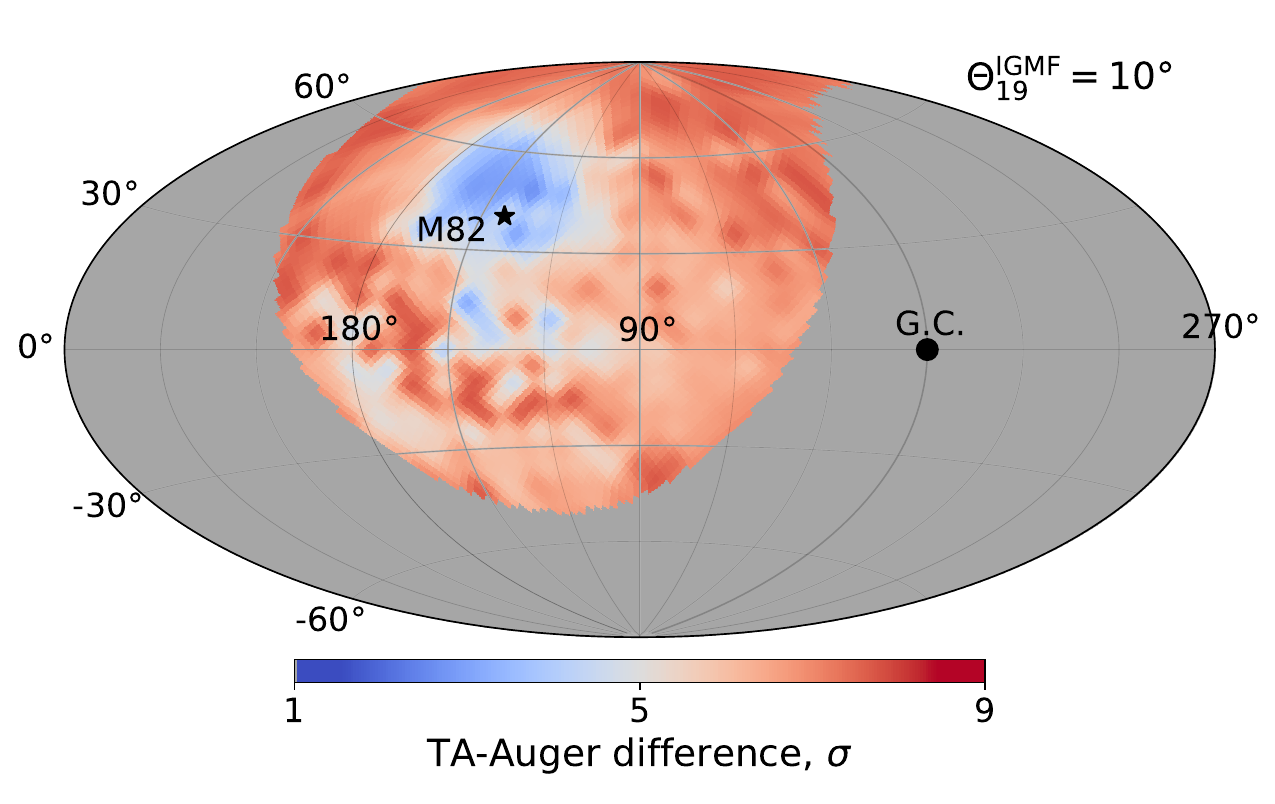}  
    \includegraphics[width=0.49\linewidth]{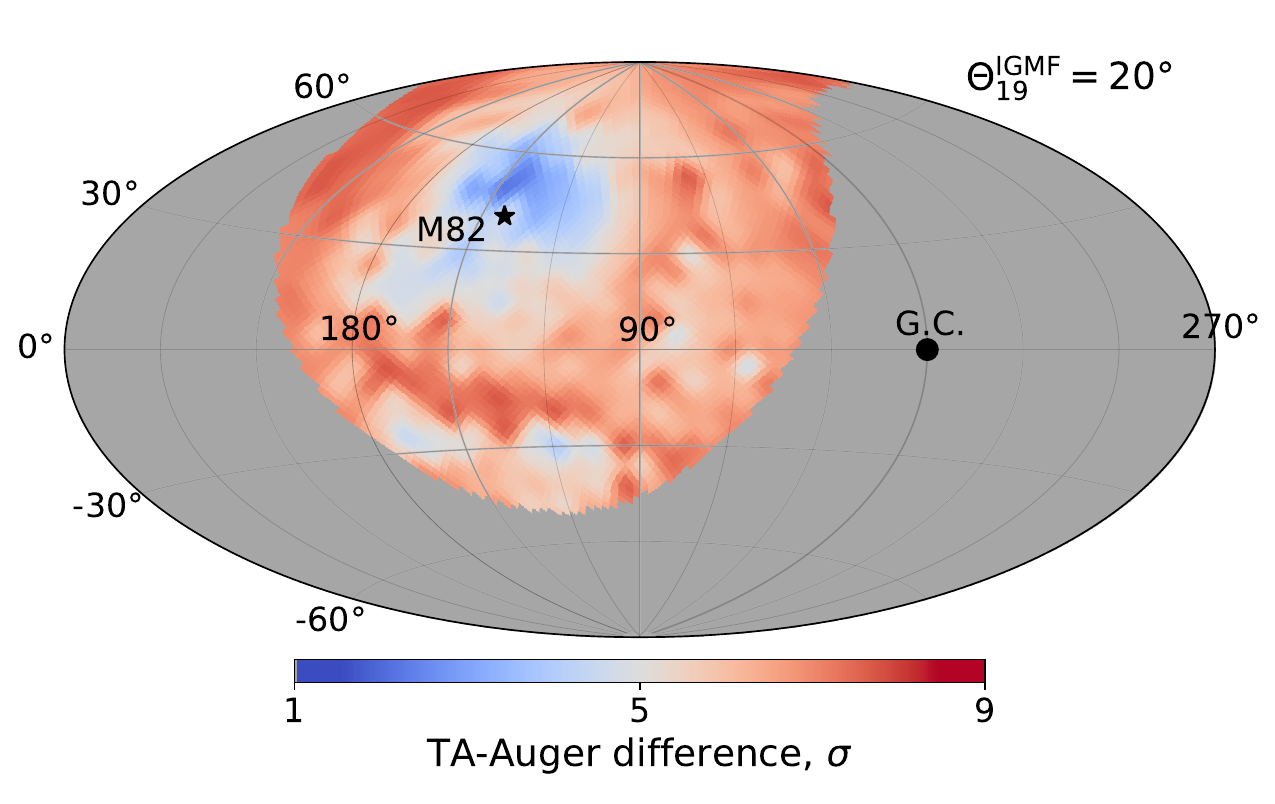}  
    \includegraphics[width=0.49\linewidth]{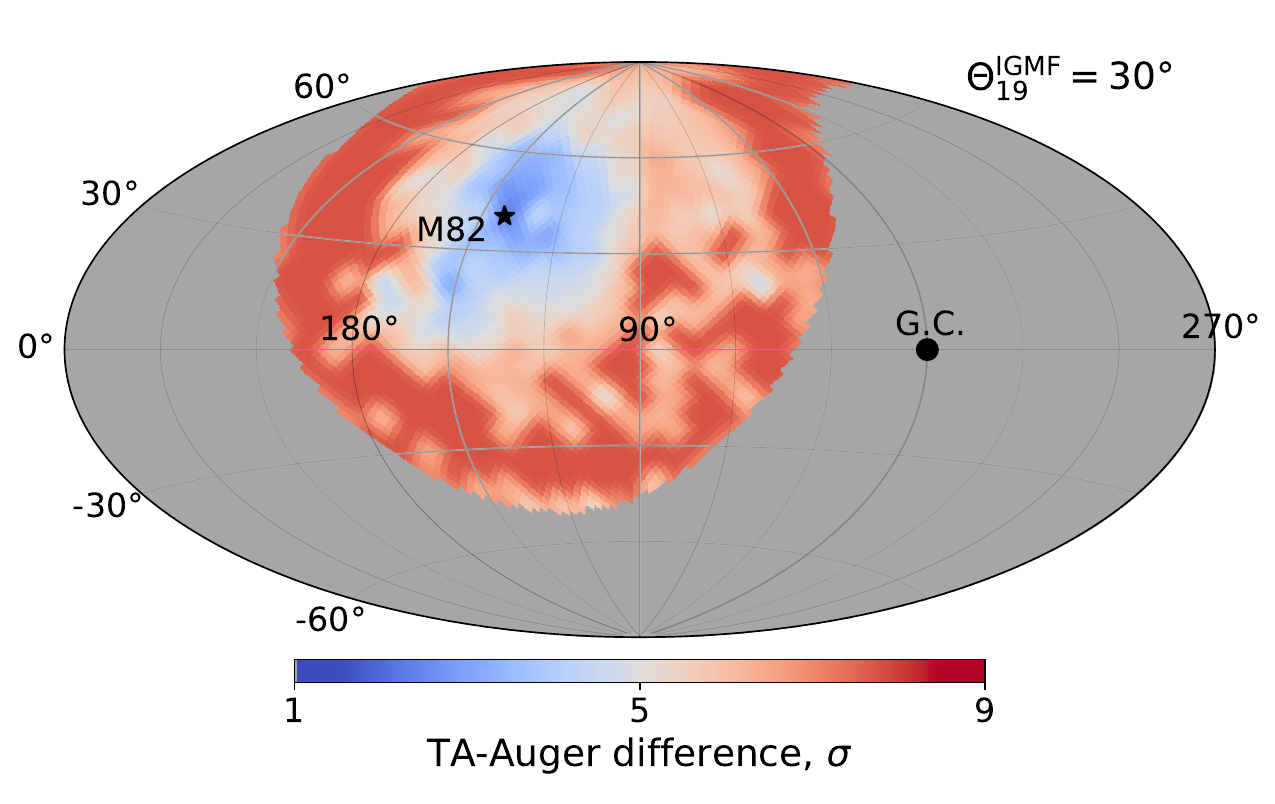}  
    \caption{Significance of the residual TA–Auger spectral discrepancy as a function of source position $(l, b)$ plotted in Galactic coordinates. Different panels correspond to different levels of additional IGMF smearing, $\theta_\mathrm{19}^\mathrm{IGMF}$. In each skymap, the Galactic Center (GC) is shifted to the right for better visibility. The blue regions on the maps correspond to the positions of the source where the TA-Auger discrepancy is significantly reduced.}
    \label{fig:source_position}
\end{figure*}

To evaluate the quality of the fits, we minimize the $\chi^2$ as described in the previous section. First, we fit both the TA and Auger spectra using only the background component $I(E,\Omega)_\mathrm{bkg}$, which includes the Auger-measured dipole, while neglecting an additional nearby source. In the energy range from $10^{18.8}$ to $10^{19.5}$ eV, the best fit yields $\chi^2/\mathrm{ndf} = 11.0/11$ for a model with one spectral break (4 parameters and 15 data bins), indicating excellent agreement between the two spectra. However, when the energy range is extended to the highest available data bins, the fit quality worsens significantly to $\chi^2/\mathrm{ndf} = 110.5/25$ for a model with two spectral breaks (6 parameters and 31 data bins). This corresponds to an $7.1\sigma$ discrepancy between the spectra, see Fig.~\ref{fig:spec_TA_Auger}. A discrepancy with an even higher significance of $8\sigma$ was reported in~\cite{TA_8sigma}, albeit using a different data and analysis.

The inclusion of the source adds six free parameters to the model ($\gamma$, $R_\mathrm{cut}$, $\beta$, $f_\mathrm{C}$, $f_\mathrm{Si}$, $f_\mathrm{Fe}$). Our fits showed that the proton $f_\mathrm{H}$ and helium $f_\mathrm{He}$ fractions are always consistent with zero, so they were fixed at zero to enhance the stability of the fit. Combined with the background parameters, this yields a total of 12 free parameters. For the combined background+source model, we scan over the source positions, fitting the parameters at each point. The significance of the remaining discrepancy as a function of the source position is shown in Fig.~\ref{fig:source_position}. The different panels correspond to different levels of additional angular smearing by the IGMF.

As shown in Fig.~\ref{fig:source_position}, there is a well-localized region in the sky where the discrepancy is significantly reduced (blue areas in Fig.~\ref{fig:source_position}). For the model without IGMF smearing, the best-fit yields $\chi^2/\mathrm{ndf}=39.6/19$, corresponding to a $2.9\sigma$ significance of the TA-Auger difference. The overall best fit is achieved for an IGMF smearing of $\theta_{19}^\mathrm{IGMF} = 20^\circ$, with $\chi^2/\mathrm{ndf} = 31.2/19$ ($2.1\sigma$). For stronger smearing, the fit quality worsens, as the source image becomes so large that it significantly contributes to the Auger spectrum regardless of its position, and  thus reducing its ability to create a difference between the TA and Auger observations.
\begin{figure*}
    \centering
    \includegraphics[width=0.49\linewidth]{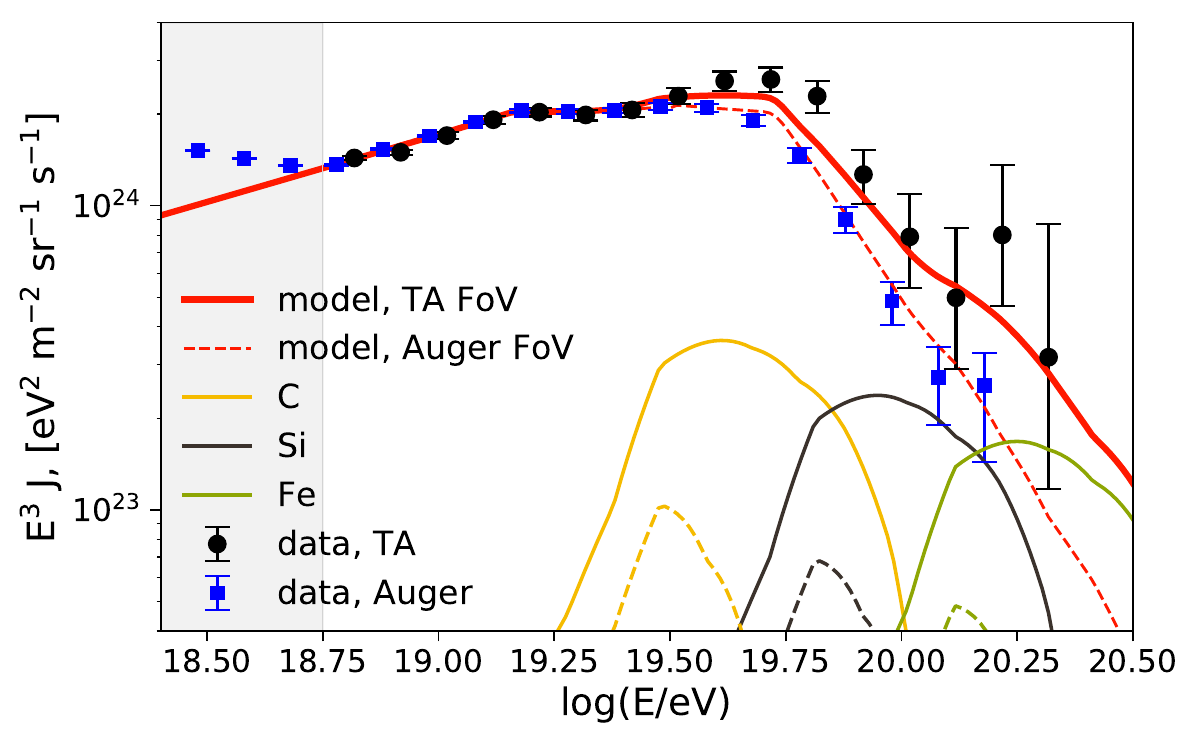}
    \includegraphics[width=0.49\linewidth]{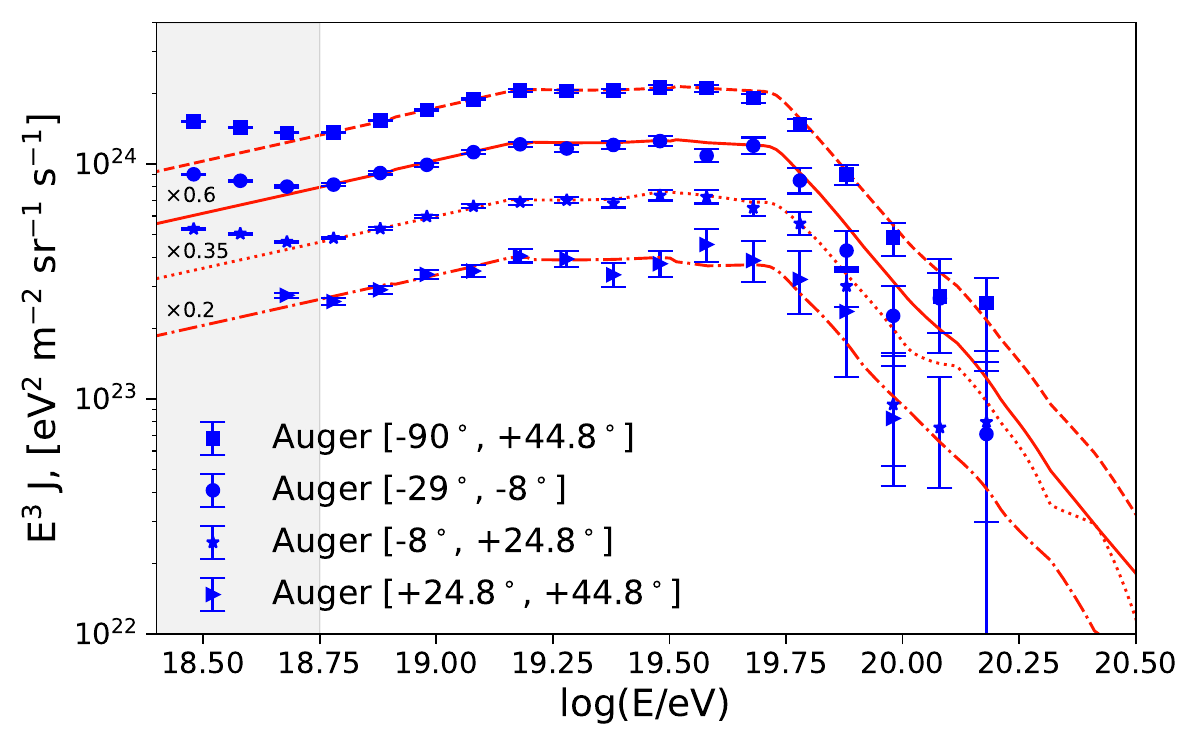}
    \caption{\textit{Left panel:} Spectral fit to the Auger and TA data, shown with dashed and solid red lines, respectively. The source contributions of C, Si, and Fe nuclei are indicated by yellow, black, and green lines, where solid lines show the source contribution to the TA spectrum and dashed lines --- to Auger spectrum. \textit{Right panel:} Auger spectrum in the three declination bands closest to the TA field of view from Ref.~\cite{PierreAuger:2025hnw}, compared with the model predictions.}
    \label{fig:excess_model_fit}
\end{figure*}

Thus, our model demonstrates that the observed TA-Auger discrepancy can be largely explained by the contribution of a nearby source without overproducing the medium-scale anisotropy. The preferred source location is stable against IGMF smearing and includes the brightest nearby starburst galaxy, M82, located at $D \approx 4$~Mpc. On the other hand, the best fit is achieved with an additional $20^\circ$ IGMF smearing. Assuming a source distance of $D = 4$ Mpc and an IGMF correlation length of $L_\mathrm{c} = 200$ kpc, this corresponds to an IGMF strength of approximately 10 nG.  Note that M82 is located in the same local sheet of the large-scale structure as the Milky Way, and existing upper limits on the cosmological magnetic field, $B < 1$ nG \cite{Durrer:2013pga,Pshirkov:2015tua}, are not directly applicable to the region between the local source and our Galaxy. 

To better illustrate how a nearby source can resolve the TA-Auger discrepancy, below we discuss in detail the best-fit configuration with zero IGMF smearing. Although we focus on this specific case, we have verified that our conclusions are general. Other configurations that provide a good fit require similar source properties and create a similar anisotropy pattern.

\subsection{Spectrum}
\label{sec:results_spectrum}
\begin{figure*}
    \centering
    \raisebox{-0.098cm}{\includegraphics[width=0.33\linewidth]{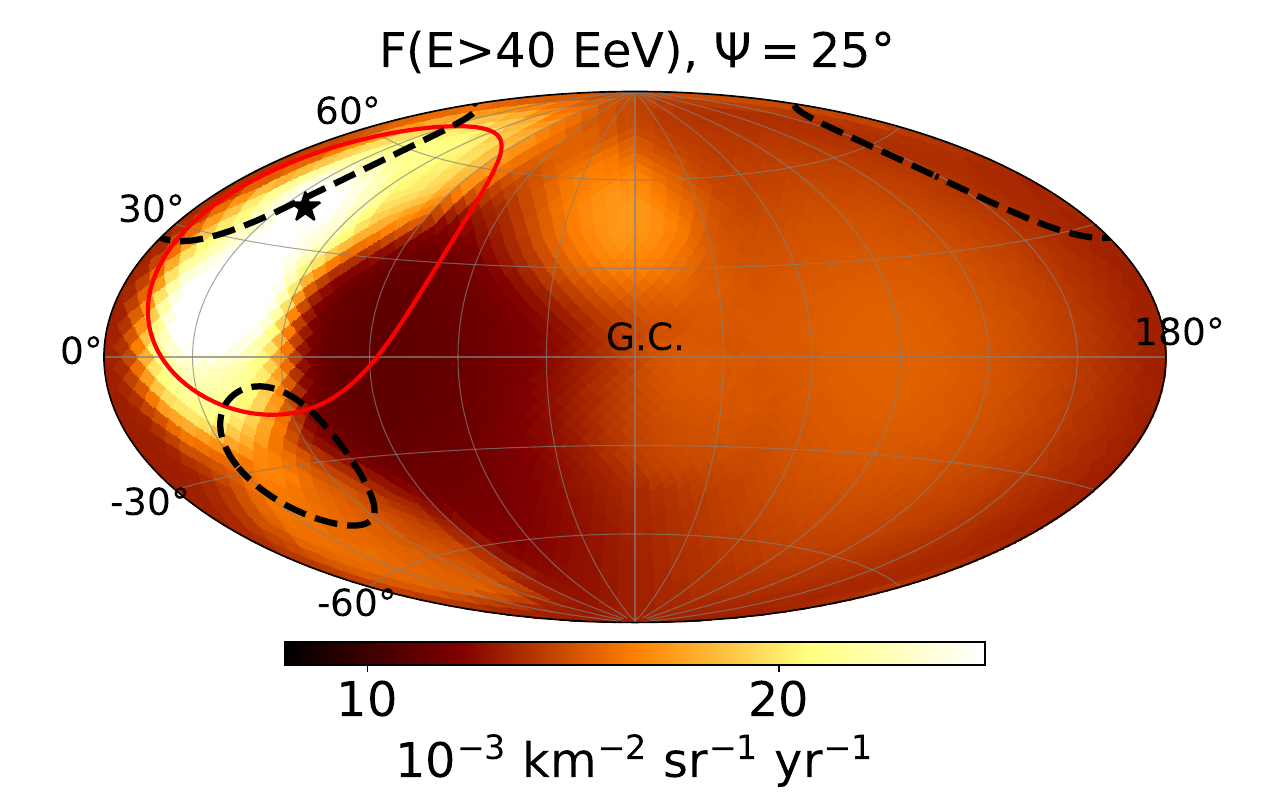}}
    \includegraphics[width=0.33\linewidth]{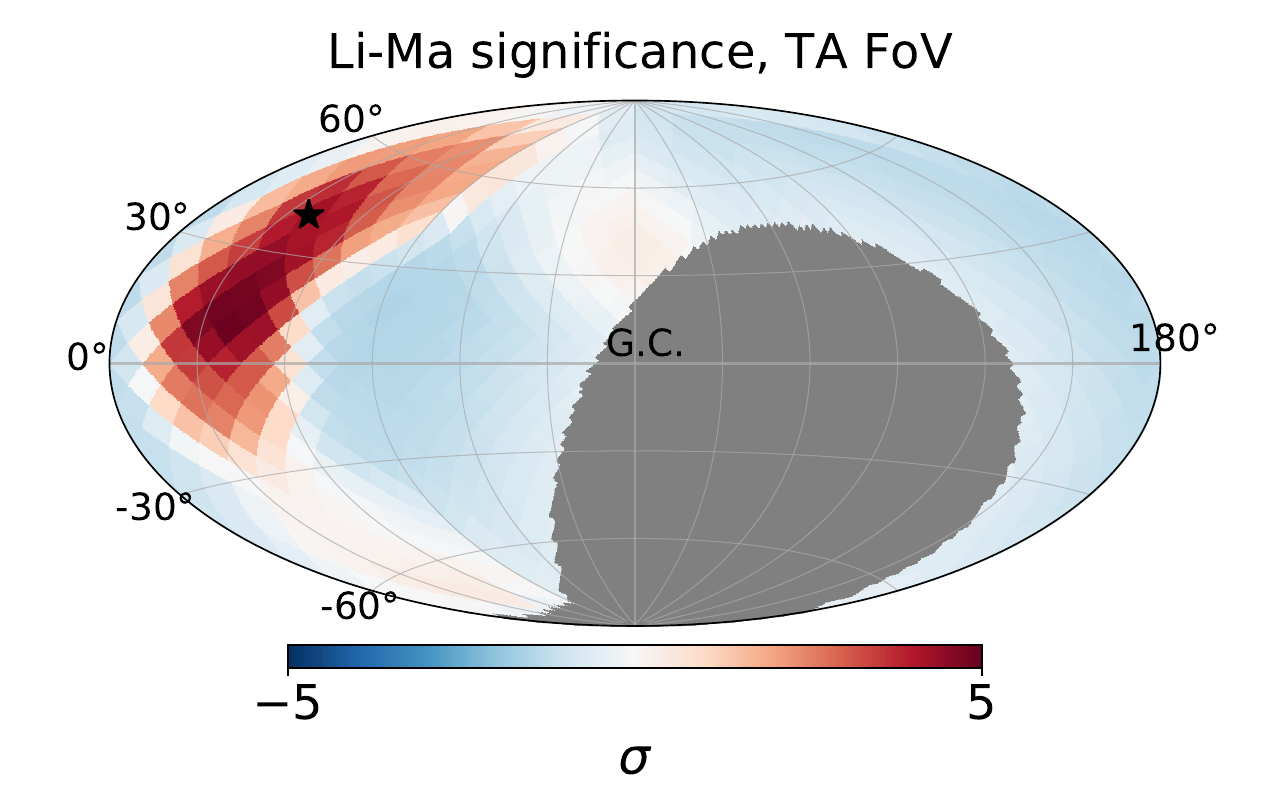}
    \includegraphics[width=0.329\linewidth]{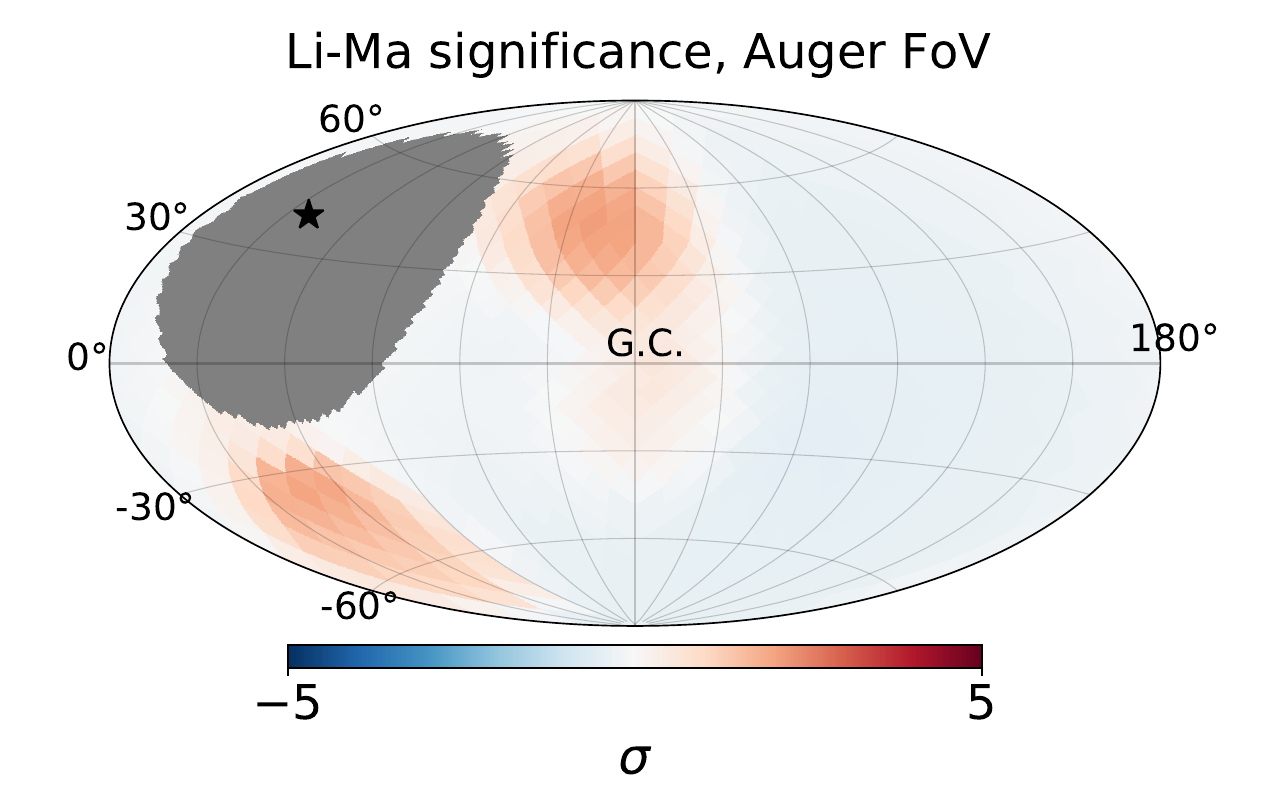}
    \caption{{\it Left panel:} UHECR flux map above 40 EeV, averaged over top-hat windows of radius $\Psi = 25^\circ$ for the best-fit model without IGMF. Black dashed circles indicate the TA hotspots, while the red contour shows the region of the sky not visible to Auger. {\it Middle panel:} Li–Ma significance map of $25^\circ$-radius hotspots above 40 EeV in the TA field of view based on the flux map from the left panel. {\it Right panel:} same as the middle panel, but for the Auger field of view. All maps are shown in Galactic coordinates, centered on the Galactic Center (G.C.). The black star indicates the source position. The gray regions in the middle and right panels indicate the portions of the sky not visible to TA and Auger, respectively.
    \label{fig:excess}}
\end{figure*}

The best-fit source position for the case of zero IGMF is $(l, b) = (141^\circ, 42^\circ)$. The contribution of this source to the TA and Auger spectra is shown in the left panel of Fig.~\ref{fig:excess_model_fit}. The source contributes significantly to the TA spectrum at $E>10^{19.5}$ eV (solid lines), while its contribution to the Auger spectrum is only mild (dashed lines). The skymap of the best-fit flux distribution is shown in the left panel of Fig.~\ref{fig:excess}. The bright yellow strip on the left side of the map corresponds to the source image, while the moderate flux enhancement on the right side results from the background dipole. Although the source can explain the major part of the TA-Auger difference, the TA data points around $10^{19.8}$~eV remain above the model predictions. This occurs because a brighter source would overproduce the medium-scale anisotropy. The central panel of Fig.~\ref{fig:excess} shows that the significance of the source image has already reached $5\sigma$ in the TA field of view. This suggests that the remaining discrepancy may be due to other sources or to an additional systematic difference between the TA and Auger measurements.

The colored solid and dashed lines in Fig.~\ref{fig:excess_model_fit} show the observed spectrum of the source as seen by TA and Auger, respectively. Due to flux amplification by the GMF, which generally depends on rigidity, this observed spectrum differs from the injection spectrum. The best-fit parameters of the injection spectrum are listed in Table~\ref{tab:src_spec}. Regardless of the IGMF smearing, the source spectrum must be hard, with a spectral index of $\gamma \approx 0$ and a cutoff rigidity of approximately $R_\mathrm{cut} \approx 10^{18.5}$~EV. In all cases, the source composition is dominated by the CNO group, with additional contributions from Si and Fe.

\begin{table}
    \begin{tabular}{l|r|r|r|r}
    \hline
    \hspace{0.1cm} parameter \hspace{0.1cm} & \hspace{0.0cm} $\theta_{19}^\mathrm{IGMF} = 0^\circ$ \hspace{0.0cm}   & \hspace{0.25cm} $10^\circ$ \hspace{0.25cm} & \hspace{0.25cm} $20^\circ$ \hspace{0.25cm} & \hspace{0.25cm} $30^\circ$ \hspace{0.25cm} \\[5pt]
    \hline
        $\log_{10}(R_\mathrm{cut}/V)$  &  $18.6$  & $18.5$ & $18.5$ & $18.6$  \\
        $\gamma$  &  $-0.8$  & $-0.6$ & $-0.1$ & $0.3$   \\
        $\beta$  &  $1.35$  & $1.17$ & $1.11$ & $1.10$   \\[5pt]
        \hline
        $f_\mathrm{C}$  &  0.86  & 0.83   & 0.88 & 0.95    \\
        $f_\mathrm{Si}$ &  0.11  & 0.13   & 0.10 & 0.02   \\ 
        $f_\mathrm{Fe}$ &  0.03  & 0.04   & 0.02 & 0.03 \\[5pt]
        \hline
        $\chi^2/\mathrm{ndf}$ & 39.6/19 & 35.7/19 & 31.2/19 & 35.0/19 \\[5pt]
        \hline
    \end{tabular}
    \caption{Best-fit parameters of the source spectrum for different levels of the additional IGMF smearing.}
    \label{tab:src_spec}
\end{table}

Surprisingly, these properties closely match those required to reproduce the Auger-measured spectrum and mass composition. Indeed, the injection spectrum parameters derived for the population of sources in, e.g., \cite{2023PhRvD.107j3045E, Bister:2023icg, 2024JCAP...01..022A, 2025arXiv250219324R}, are very similar to those listed in Table~\ref{tab:src_spec}. This is particularly interesting since recently it was shown that a population of nearly identical sources (``standard candles'') is needed to fit the Auger composition data~\cite{2023PhRvD.107j3045E}. This suggests that the source responsible for the TA-Auger discrepancy may simply be the nearest representative of such a population of standard-candle sources.

\begin{figure*}
    \centering
    \includegraphics[width=0.33\linewidth]{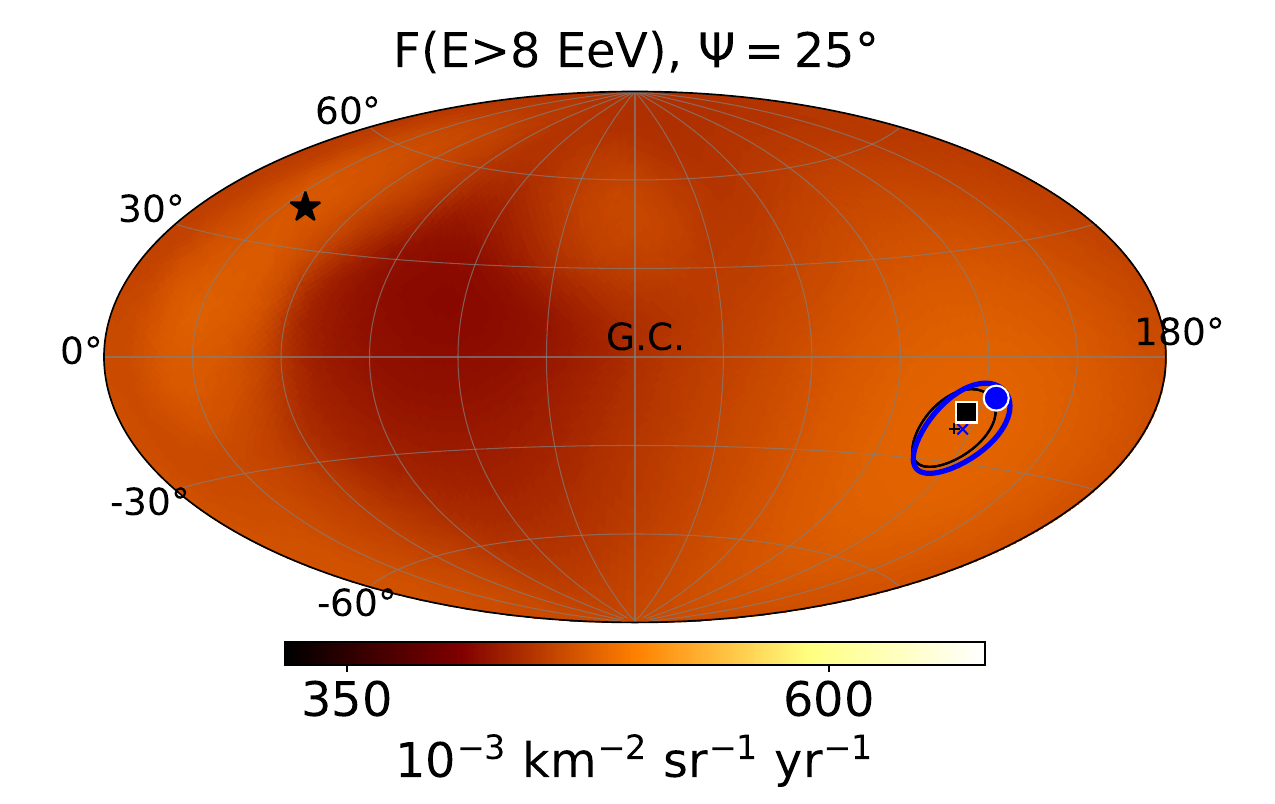}
    \includegraphics[width=0.33\linewidth]{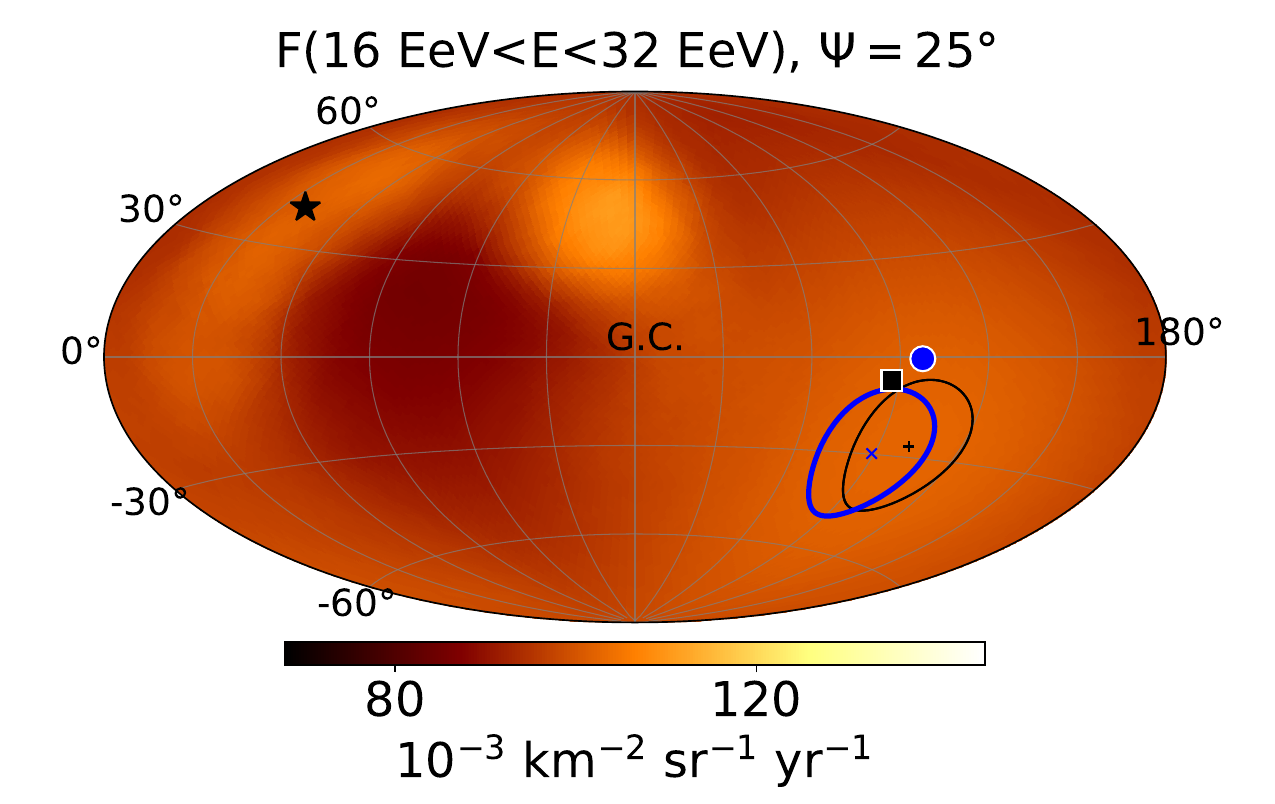}
    \includegraphics[width=0.329\linewidth]{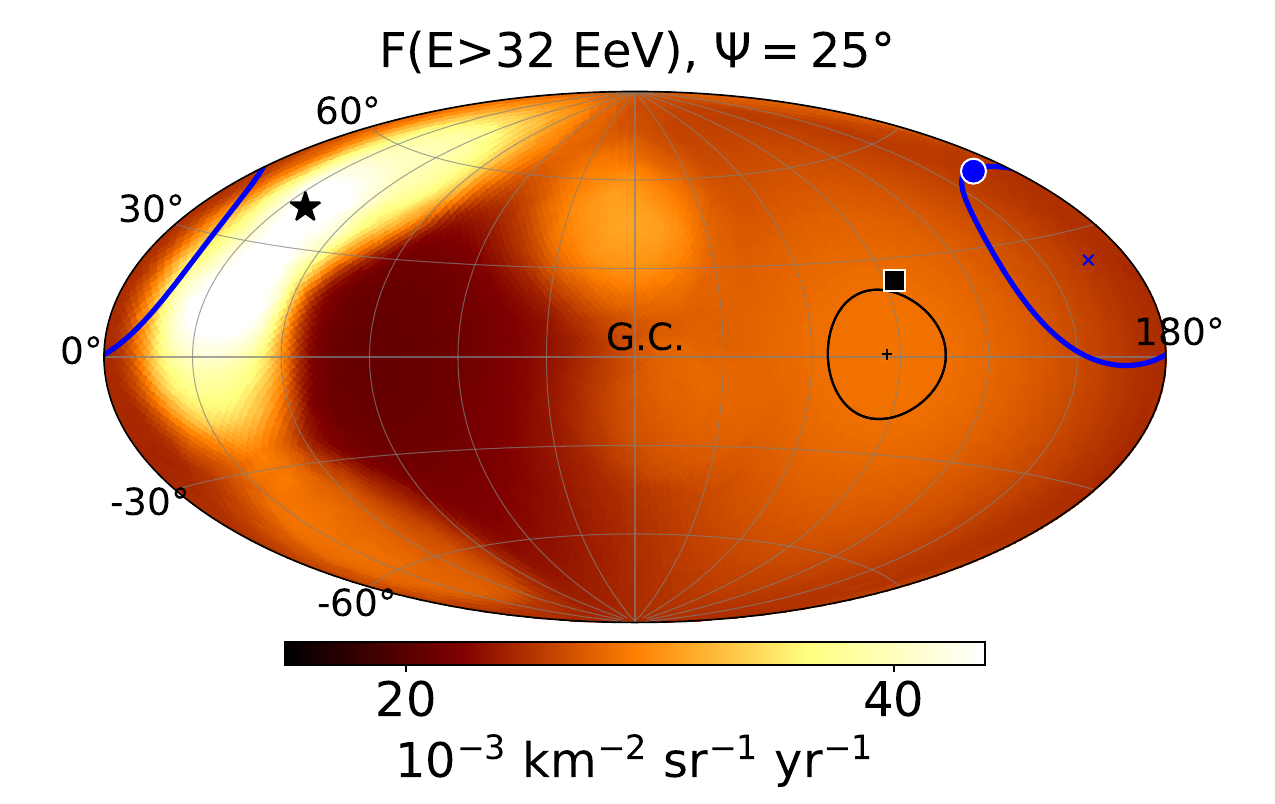}
    \caption{Dipole direction at different energies. Black cross with uncertainty contours show the dipole direction from Auger-only data, while blue cross and contour shows the dipole direction from the full sky analysis~\cite{TelescopeArray:2025yvu}. Black square and blue dot indicate the dipole direction in the best-fit model (background with Auger-measured dipole plus source, zero IGMF), calculated using only the Auger field of view and the full sky, respectively. The black star marks the source position. The background in each panel shows the best-fit model flux map for the corresponding energy range.
    \label{fig:dipole}}
\end{figure*}

Although the source is mainly visible to TA, it also contributes to Auger and slightly modifies its full-sky spectrum, as shown in Fig.~\ref{fig:excess_model_fit}. This contribution is also clearly visible in the right panel of Fig.~\ref{fig:excess}, which shows two mildly significant hotspots near the edges of the Auger field of view created by the source. In addition to hotspot analysis, a more straightforward test has recently become possible with the Auger collaboration's release of the UHECR spectrum for five declination bands~\cite{PierreAuger:2025hnw}. We tested our model against all five spectra and found good agreement. The source contribution is small not only in the full Auger sky but also in each individual declination band, as shown in the right panel of Fig.~\ref{fig:excess_model_fit}. Thus, we directly demonstrate that there is no contradiction between non-observation of UHECR flux declination dependence by Auger and TA-Auger difference.

\subsection{Dipole}
\label{sec:results_dipole}
  
Apart from the spectral difference between TA and Auger, the dipole direction at the highest energies ($E > 32$~EeV) shifts by approximately $90^\circ$ when reconstructed from the combined Auger-TA dataset, compared to the Auger-only data~\cite{TelescopeArray:2025yvu}. Interestingly, the shift occurs exactly at those energies at which the spectral bump appears in the TA data, suggesting a common origin. Below, we demonstrate that the best-fit source found by spectral fitting also produces the observed shift in the dipole direction.

As discussed in Section~\ref{sec:setup}, we assume a background with fixed dipolar anisotropy as measured by Auger. The corresponding dipole directions are shown with black contours in Fig.~\ref{fig:dipole}, where different panels correspond to different energy ranges. To estimate the effect of the source on the dipole direction, we performed two tests using the UHECR flux map with the best-fit source.  First, we calculated the predicted dipole direction excluding the portion of the sky not visible to Auger. Second, we repeated the calculation using the full sky.

In Fig.~\ref{fig:dipole}, the predicted dipole directions are shown with a black square and blue dot for the Auger field of view and full sky, respectively. As expected, at low energies ($E>8$ EeV), the inclusion of the source has only a minor effect on the dipole: in all cases, the direction remains close to the initial one, marked by the black cross. However, the effect of the source becomes more pronounced at higher energies.

In the energy range $16$–$32$ EeV, the source contribution shifts the dipole direction by approximately $20^\circ$ in both the Auger-only and full-sky analyses, though the two directions remain close to each other. Finally, at the highest energies, the source produces a significant shift in the dipole direction for the full-sky field of view, bringing it into agreement with the results of the combined Auger+TA analysis. Thus, in our toy model, the bright source does not spoil the consistency of the dipole directions at low energies while simultaneously explaining the discrepancy between the dipole measurements reported by Auger and Auger+TA at $E>32$~EeV.

\subsection{Hotspots}
\label{sec:results:hotspots}

Our model is constructed so that it does not produce localized anisotropies that would be detected with more than $5\sigma$ significance by TA or $2.5\sigma$ by Auger. It does, however, generate a broad, lower-significance excess --- mainly in the TA field of view --- that extends from the source position to the Galactic plane and beyond (see Fig.~\ref{fig:excess}, left panel). This geometry of the excess arises from the combined effect of random deflections and the strong magnetic field in the outer Galaxy in the \texttt{KST24} model.

As mentioned in the Introduction, both TA and Auger have reported several localized excesses on medium angular scales: the TA hotspot at $(l, b) = (177^\circ, 50^\circ)$~\cite{TA_hotspot}; the TA ``warm spot'' in the direction of the Perseus–Pisces supercluster, $(l, b) = (129^\circ, -28^\circ)$~\cite{TA_warmspot}; and the Auger Centaurus A excess at $(l, b) = (305^\circ, 16^\circ)$~\cite{PierreAuger:2010ofq,PierreAuger:2024hrj}. Earlier evidence for medium-scale anisotropy was reported by the authors of \cite{Kachelriess:2005uf} towards $(l, b) = (140^\circ, 70^\circ)$. Several attempts have been made to associate these medium-scale anisotropies with potential UHECR sources, taking into account deflections in the Galactic and extragalactic  magnetic fields~\cite{Soriano:2019wfx,Allard:2023uuk,Dolgikh:2025bac}. 

Interestingly, both TA hotspots partially overlap with the broad excess predicted by our model. One may speculate that both hotspots originate from this broad excess, with the apparent mismatch in morphology explained by limited statistics. Since most of the excess lies within the TA field of view, this would be consistent with the negative result of the TA hotspots cross-check by Auger~\cite{PierreAuger:2024hrj}. 

Due to deflections in the magnetic field, the cosmic-ray flux from the source also “leaks” into the Auger field of view, producing a marginal excess near the edge of the Auger exposure, too far from the Cen A region to suggest any physical connection between the two (for the recent analysis of the Cen A region excess, see~\cite{Bister:2025emc}). It coincides, however, with a mild excess (or upward fluctuation) in the Auger data \cite{2024ApJ...976...48A}, so it will be difficult to rule out even with increased statistics.

\subsection{Amaterasu and Oh-my-God particles}
\label{sec:results_amaterasu}

\begin{figure}
    \centering
    \includegraphics[width=0.9\linewidth]{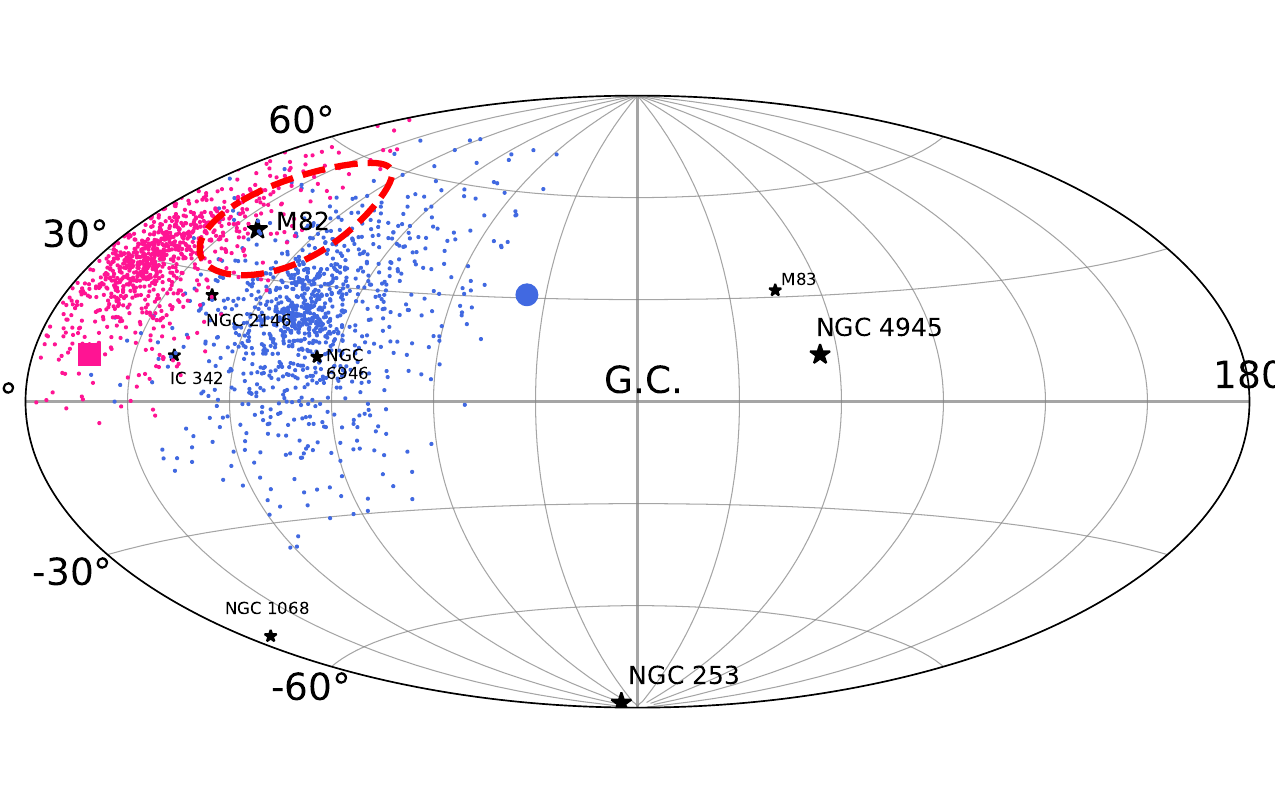}
    \caption{    Potential source locations of the Amaterasu and Oh-My-God particles. The large blue and red dots indicate the arrival directions of the Amaterasu and Oh-My-God particles, respectively. The small dots show the backtracked directions in the \texttt{KST24} GMF model for $N = 1000$ realizations of the turbulent magnetic field, including additional IGMF smearing of $\theta^\mathrm{IGMF}_{19} = 20^\circ$. }
    \label{fig:amaterasu_ohmygod}
\end{figure}

In the previous sections, we have identified a well-defined region of the sky as a plausible source location. Meanwhile, observation of two highest-energy cosmic ray events --- the Oh-My-God particle with an energy of $320\pm90$ EeV detected by HiRes~\cite{HIRES:1994ijd} and the Amaterasu particle with ($E=244\pm29\mathrm{\,(stat.)^{+51}_{-76}\mathrm{\,(syst.)}}$~EeV) detected by TA~\cite{TelescopeArray:2023sbd} --- suggests the existence of nearby sources~\cite{2024JCAP...04..042K}, as the propagation distance of cosmic rays of highest energy is limited by energy losses. Nevertheless, these sources have not yet been firmly identified by previous 
backtracking analyses~\cite{2024JCAP...04..042K, Unger:2023hnu, Bourriche:2024bbe, Korochkin:2025ugg}.

It is therefore interesting to check whether the source region identified in our analysis is consistent with the backtracked directions of the Oh-My-God and Amaterasu particles, which we assume to be iron nuclei. If no additional smearing is applied ($\theta^\mathrm{IGMF}_{19} = 0^\circ$), the backtracked directions of these events lie near the favored region but do not overlap with it, in agreement with previous studies. However, when the smearing is set to $\theta^\mathrm{IGMF}_{19} = 20^\circ$, as favored by our fits, the backtracked directions of both particles overlap with the favored region (see Fig.~\ref{fig:amaterasu_ohmygod}), suggesting that they may originate from the same source responsible for the Auger-TA spectral difference.

\section{Discussion and Summary}
\label{sec:summary}

The different locations of the Auger and TA experiments in the Southern and Northern hemispheres, respectively, and consequently their different fields of view, provide a natural explanation for the observed differences in their energy spectra at the highest energies. However, a large fluctuation in the source distribution is required to make the Northern sky sufficiently different from the Southern one. Such a fluctuation could arise if one or a few bright sources were anomalously close. Since no bright isolated sources have been detected, the question arises whether this scenario is viable.

We have demonstrated that even a single source, superimposed on a smooth and nearly uniform background, can be smeared and deflected by the Galactic (and possibly extragalactic) magnetic fields in such a way that it contributes mostly to the TA field of view, producing the spectral difference without overproducing medium-scale anisotropy. To achieve this, the source must be rather special: it should have a hard spectrum in order not to spoil the Auger–TA agreement below $10^{19.5}$~eV, and an intermediate-mass composition to produce the right magnitude of deflections. Interestingly, sources with similar properties have been proposed to fit the Auger-only spectrum and composition \cite{PierreAuger:2025lye}. If the TA indication of a heavier composition at the highest energies \cite{TelescopeArray:2024oux,TelescopeArray:2024buq} is confirmed, one may be able to account for all observations with a single population of sources. 

This scenario turns out to have a number of additional advantages. First, when deflected in the \texttt{KST24} magnetic field and smeared by random deflections, the source produces a broad, mild excess that partially overlaps with both TA hotspots, suggesting their possible explanation. Second, the excess of cosmic-ray flux in the Northern hemisphere at the highest energies naturally explains the sudden $\sim 90^\circ$ shift in the dipole direction observed in the combined Auger–TA data compared to the Auger-only analysis. This change occurs only at the highest energies, while at lower energies both the Auger–TA and Auger-only analyses yield consistent dipole directions.

Finally, the best-fit position of the source lies close to M82, the brightest nearby starburst galaxy (see Fig.~\ref{fig:source_position}). This galaxy was suggested as a significant contributor to the UHECR flux well before the TA hotspots were discovered and a heavy-mass composition was proposed by \cite{Anchordoqui:2001ss}, and was later discussed among other possible sources in Ref.~\cite{He:2014mqa}. Starburst galaxies have also been argued to be likely hosts of UHECR sources in Ref.~\cite{Marafico:2024qgh}. The required extragalactic deflections found there are consistent with our preferred value of $\sim 20^\circ$, corresponding to a magnetic field of about $10$~nG in the local filament. Interestingly, M82 could also be a possible source of the highest-energy Amaterasu and Oh-My-God particles (see Fig. \ref{fig:amaterasu_ohmygod}) if one assumes such extragalactic fields.

\vskip 0.cm

\section*{Acknowledgments}
\noindent The work of DS has been supported in part by the French National Research Agency (ANR) grant ANR-24-CE31-4686. The work of AK and PT is supported by the IISN project No. 4.4501.18.

\bibliographystyle{apsrev4-2}
\bibliography{refs}

\end{document}